\documentclass{desyprocA4}

\usepackage{amssymb}
\usepackage{amsmath}
\usepackage{multirow}
\usepackage[english]{babel}
\usepackage[]{graphicx}
\usepackage{slashed}
\usepackage{hep,hepnicenames}

\newcommand{\eeAl}{\APelectron\Pelectron \to \hzero\Azero}

\newcommand{\eeAlH}{\APelectron\Pelectron \to \hzero\Azero/\Hzero\Azero}
\newcommand{\eeZl}{\APelectron\Pelectron \to \hzero\PZ^0}

\newcommand{\eeZlH}{\APelectron\Pelectron \to \hzero\PZ^0/\Hzero\PZ^0}

\newcommand{\retildehat}{\ensuremath{\mbox{Re}\,\hat{\Sigma}}}

\newcommand{\CP}{\ensuremath{CP}}
\newcommand{\hzero}{\ensuremath{\PHiggslightzero}} 
\newcommand{\Hzero}{\ensuremath{\PHiggsheavyzero}} 
\newcommand{\Azero}{\ensuremath{\PHiggspszero}} 

\begin{document}
\title{Higgs boson production at Linear Colliders from a generic 2HDM: 
the role of triple Higgs self-interactions}

\author{{\slshape David L\'opez-Val$^1$, Joan Sol\`a$^2$} \\[1ex]
$^1$Institut f\"ur Theoretische Physik, Universit\"at Heidelberg \\
Philosophenweg 16, 67119 Heidelberg, Germany. \\ 
\\[1ex]
$^2$High Energy Physics Group, Dept. ECM, and Institut de Ci{\`e}ncies del Cosmos\\
Universitat de Barcelona \\ Av. Diagonal 647, E-08028 Barcelona, Catalonia, Spain.}


\maketitle

\begin{abstract}

We review selected results for Higgs boson production at Linear Colliders
in the framework of the general Two-Higgs-Doublet Model (2HDM). We concentrate
on the analysis of i) the pairwise production of neutral Higgs bosons ($\hzero\Azero,\Hzero\Azero$);
and ii) the neutral Higgs boson-strahlung modes ($\hzero\PZ^0$, $\Hzero\PZ^0$). We identify
sizable production rates, in the range of $\sigma \sim \mathcal{O}(10-100)$ fb for $\sqrt{s} = 0.5$ TeV,
alongside with large quantum effects ($\delta_r\sim \pm 50\%$), which we can 
fundamentally track down to the enhancement power
of the triple-Higgs self-interactions. This constitutes
a telltale signature of the 2HDM, with no counterpart in e.g. 
the Minimal Supersymmetric Standard Model (MSSM). We compare these results with several complementary
double and triple $\mathcal{O}(\alpha^3_{ew},\alpha^4_{ew})$ Higgs-boson production mechanisms
and spotlight a characteristic phenomenological profile which could eventually be 
highly distinctive of a non-supersymmetric two-Higgs-doublet structure.

\end{abstract}

\section{Introduction}

Deciphering the fundamental nature of Electroweak Symmetry Breaking (EWSB)
lies at the very frontline of the current theoretical and experimental
research in Particle Physics. Even in spite of the tantalizing Higgs boson candidates recently identified
by the ATLAS and CMS experiments \cite{evidence}, a long way might yet stand ahead of us
until we are able to convincingly close in on such a longstanding conundrum.
In particular, were these signatures finally confirmed, a first question
to be answered would be whether they can be described within the
minimal framework of the Standard Model (SM) or, on the contrary, if
they should rather be attributed to an extended EWSB sector \cite{haberrev}.   
One canonical example of the latter is the general Two-Higgs-Doublet Model (2HDM) \cite{2hdmrev}.
The model is built upon a second scalar
$SU_L(2)$ doublet with $Y=+1$ weak hypercharge. This results into a larger Higgs boson spectrum
of five physical Higgs fields: neutral \CP-even ($\hzero,\Hzero$), neutral \CP-odd ($\Azero$) 
and charged $\PHiggs^{\pm}$. 
Such a simple, but yet non-minimal extension of the SM Higgs sector 
has gathered growing attention over the past years \cite{recent2hdm} and become a 
cherised setup for  
model builders and phenomenologists alike. Besides the many novel,
and usually highly distinctive features put forward by the model
in multifarious domains -- from collider to flavor physics or 
astrophysics --,  
the 2HDM provides
a suitable low-energy description to many UV completions of the
the EWSB dynamics.

\begin{figure}[tb!]
\begin{center}
 \includegraphics[scale=0.8]{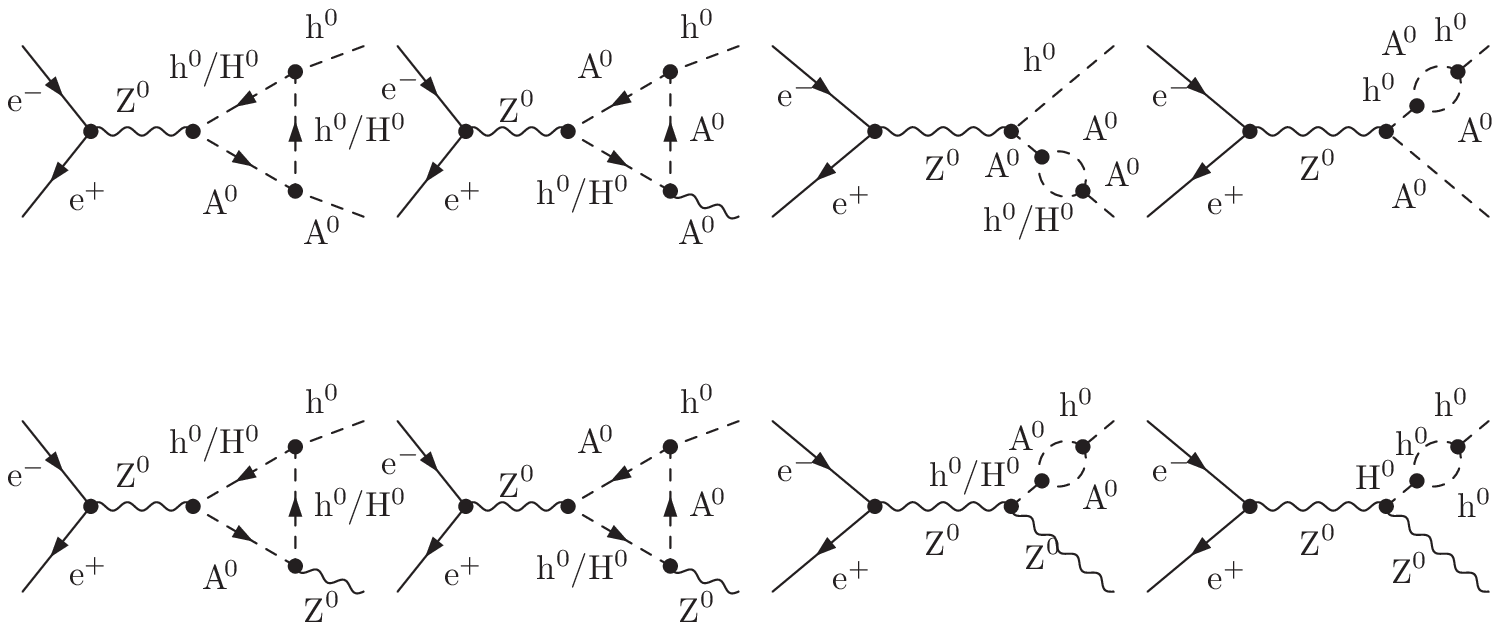}
\caption{Sample of one-loop Higgs-mediated Feynman diagrams which account 
for the bulk of the quantum effects to the
neutral Higgs pair $\APelectron\Pelectron \to \hzero\Azero$ (upper row)
and the Higgs-strahlung mechanisms $\APelectron\Pelectron \to \hzero\PZ^0$ (lower row).} \label{fig:diags}
\end{center}
\end{figure}

The model is fully specified once we fix i) the masses of the physical
Higgs bosons; ii) the ratio $\tan\beta \equiv \langle
H_2^0\rangle/\langle H_1^0\rangle$ of the two Vacuum Expectation Values (VEVs) giving masses to
the up- and down-like quarks; iii) the mixing angle $\alpha$ between the
two $\CP$-even states; and iv) one remaining Higgs boson
self-couplingin the Higgs potential, hereafter dubbed $\lambda_5$. 
We note in passing that the Higgs sector of the MSSM  \cite{mssm}
corresponds to a particular realization of the general (unconstrained) 2HDM,
for which the invariance under SUSY transformations enforces a number of additional restrictions
-- most significantly, the Higgs boson self-interactions
become tied to the gauge couplings. This situation is remarkably different in the general
2HDM, where the size of these self-interactions is fundamentally unrestricted and 
it only becomes limited, in practice, by the interplay of theoretical consistency
conditions (unitarity \cite{unitarity}, vacuum stability \cite{vacuum}) and experimental
bounds (viz. the excluded Higgs boson mass ranges from the 
from direct collider searches, and also the constraints derived from electroweak \cite{deltarho}
and flavor physics observables \cite{superiso}). A detailed account of these restrictions 
and of the model setup can be found in 
Ref.\,\cite{loop1}. For comprehensive analyses of the
2HDM parameter space constraints, see e.g. Refs.~\cite{constraints_general}.

Following the eventual discovery of the Higgs boson(s) at the LHC,
it will be crucial to address the precise experimental determination
of the corresponding quantum numbers, mass spectrum and couplings to other particles.
A linear collider (linac) can play a central role in this enterprise \cite{ILCPhysics}. 
Dedicated studies have exhaustively scrutinized the phenomenological
imprints of the basic 2HDM Higgs boson production modes,
such as e.g.
i) triple Higgs, $\APelectron\Pelectron \to 3h$ \cite{giancarlo};
ii) inclusive Higgs-pair through EW gauge boson fusion, $\APelectron\Pelectron \to V^*V^* \to 2h+X$ \cite{neil};
iii)  exclusive Higgs-pair $\APelectron\Pelectron  \to 2h$ \cite{loop1,hw};
and iv) Higgs strahlung, or associated production with a weak gauge boson $\APelectron\Pelectron  \to hV$ \cite{loop2}, 
with [$h \equiv \hzero,\Azero,\Hzero,\PHiggs^{\pm}$] and [$V  \equiv \PZ^0,\PW^{\pm}$]\footnote{For related work in the context of
MSSM Higgs boson production see e.g. \cite{mssmloop}.}. 
As a common highlight, all these studies report on
sizable production rates and large quantum effects,
arising from the potentially enhanced Higgs self-interactions. 
Interestingly enough, Higgs boson searches at $\APelectron\Pelectron$ 
colliders may also benefit from alternative running modes, particularly from $\Pphoton\Pphoton$
scattering. Processes such as $\gamma\gamma$-induced production of single ($\gamma\gamma \to h$) and double ($\gamma\gamma \to 2h$) Higgs bosons
have been studied from this viewpoint. 
These entirely operate at the quantum level, via
an effective (loop-induced) Higgs/photon interaction $g_{\Pphoton\Pphoton h}$ that we may regard
as a direct
probe of non-standard (charged) degrees of freedom coupled to the Higgs sector.
The aforementioned single Higgs channels have been considered in the framework of the SM \cite{photon_sm},
the 2HDM \cite{photon_2hdm} and the MSSM \cite{photon_mssm, previousmssm} and are known to exhibit excellent
experimental prospects, not only due to the clean environment inherent to a linac machine,
but also owing to the high attainable $\gamma\gamma$ luminosity, and the possibility to 
tune the $\Pphoton$-beam polarization as a strategy
to enlarge the signal-versus-background ratios\footnote{Analogue studies for the $\gamma\gamma\to hh$ mode
are available e.g. in Ref.~\cite{doublephoton}.}. 

\section{Numerical analysis}
\label{sec:analysis}

\begin{figure}[t!]
\begin{center}
 \includegraphics[scale=0.4]{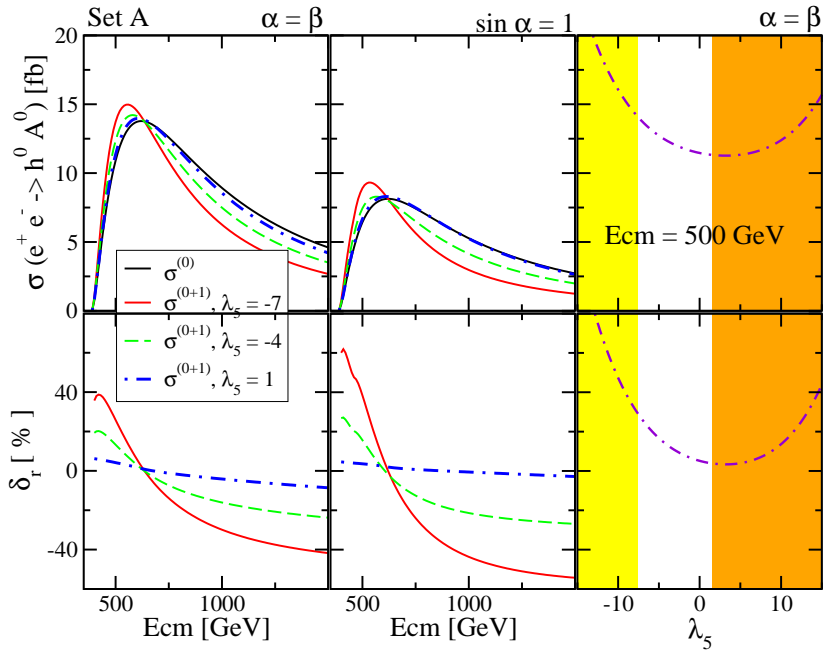}
\caption{Tree-level and loop-corrected cross section [$\sigma(\eeAl)$] (in fb),
alongside with 
the relative size of the radiative corrections 
[$\delta_r \equiv [\sigma^{(0+1)}-\sigma^{(0)}]/\sigma^{0}$] (in \%),
as a function of $\sqrt{s}$ (left, center) and $\lambda_5$ (right). We fix $\tan\beta =
1.2$ (compatible with the lower $\tan\beta$ bound from $B_d^0 -
\bar{B}_d^0$ data\,\cite{superiso}) and examine the representative
choices $\alpha=\beta$ (maximum $g_{\hzero\Azero\PZ^0}$ tree-level coupling) and
$\alpha=\pi/2$ (fermiophobic limit for $\hzero$ in type-I 2HDM). 
The influence of the Higgs self-interactions
is assessed by dialing the value of the parameter $\lambda_5$.
The shaded areas on the left (resp. right) are
excluded by unitarity (resp. vacuum stability).
} \label{fig:one}
\end{center}
\end{figure}

In this contribution we review two particular 2HDM Higgs boson production modes
at a linear collider, to wit: 
i) the pairwise production of neutral Higgs bosons $\APelectron\Pelectron \to \hzero\Azero/\Hzero\Azero$;
and ii) the associated production of a neutral Higgs and a $\PZ^0$ bosons, 
$\APelectron\Pelectron \to \hzero\PZ^0/\Hzero\PZ^0$ -- the so-called 
Higgs-strahlung mechanism, which we can regard as the 2HDM analog(s)
to the Bjorken process in the SM\,\cite{Bjorken76}. 
The motivation herewith is threefold: i) a first focus point is
to seek for the most favorable regions across the 2HDM parameter space, for which
the Higgs boson production rates become optimal, and to correlate them to alternative
multi-Higgs production modes; ii) second, we aim at quantifying the importance of the radiative
corrections associated to these processes; iii) and third, we shall examine
the impact of the Higgs boson self interactions and their potentially
enhanced strenght. The leading order production rates merely depend
on the Higgs couplings to the Z boson. 
In other words, they are
entirely subdued by the gauge symmetry -- and hence they 
feature no disclosing scenarios between the general 2HDM and e.g. the MSSM.
The resulting phenomenological portray, however, may clearly depart
once the quantum effects to such couplings are considered. 
Vertex corrections, in particular, turn out to be
sensitive to the triple Higgs self-interactions through the interchange
of virtual Higgs bosons which are then linked to the external Higgs boson legs.
A sample of such Higgs-mediated one-loop diagrams is displayed in Fig.~\ref{fig:diags}.
These effects we can roughly describe by a loop-induced form factor, which spells out
how the strength of the bare Higgs-to-gauge boson couplings is modified: 

\begin{equation}
 g_{h\Azero\PZ^0} \to   g_{h\Azero\PZ^0}\, \left[ 1 + 
\frac{|\lambda_{HHH}|^2}{16\pi^2\,s}\,f(M^2_{h}/s,M^2_{\Azero}/s) \right]
\label{eq:formfactor}.
\end{equation}

\noindent Here $\lambda_{HHH}$ stands for generic triple Higgs
self-interaction, and $1/16\pi^2$ for the standard loop integral suppression
factor. By $f(M^2_{h}/s,M^2_{\Azero}/s)$ we denote a generic rational function involving
the ratios of the different mass scales taking part in the process. The above
expression~\eqref{eq:formfactor} indicates how the Higgs-to-gauge boson
couplings, which are entirely anchored by the gauge symmetry at the leading-order,
may be strongly promoted at one-loop through the
indirect effect of the Higgs boson self-couplings -- unlike their MSSM counterparts. 


\subsection{Calculational setup}

\begin{figure}[t!]
\begin{center}
\begin{tabular}{ccc}
 \includegraphics[scale=0.85]{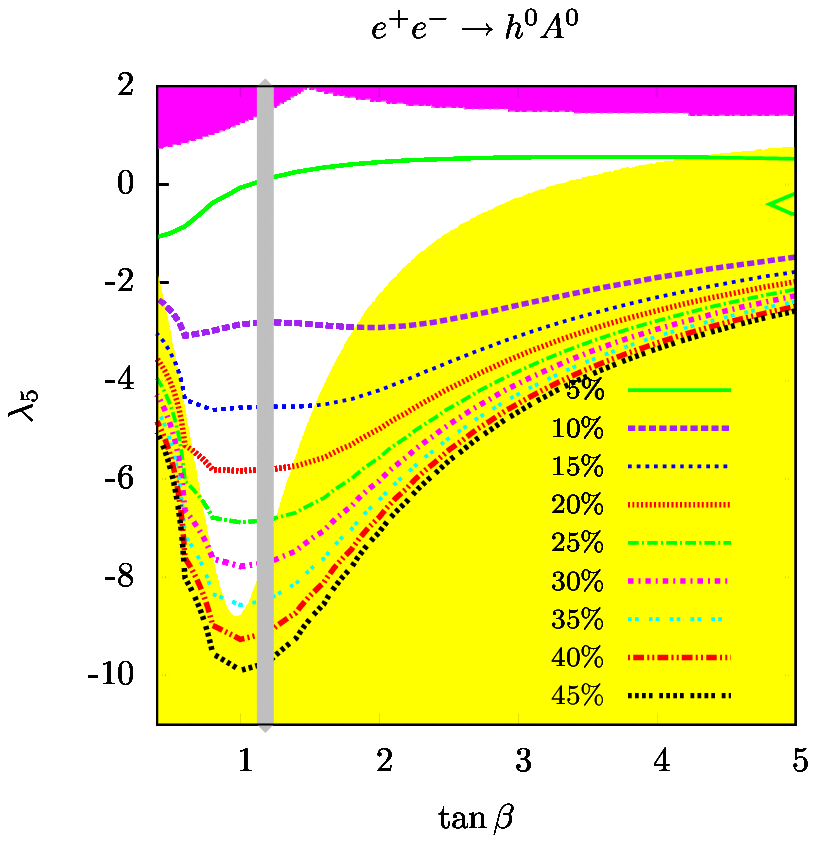} & & 
\includegraphics[scale=0.85]{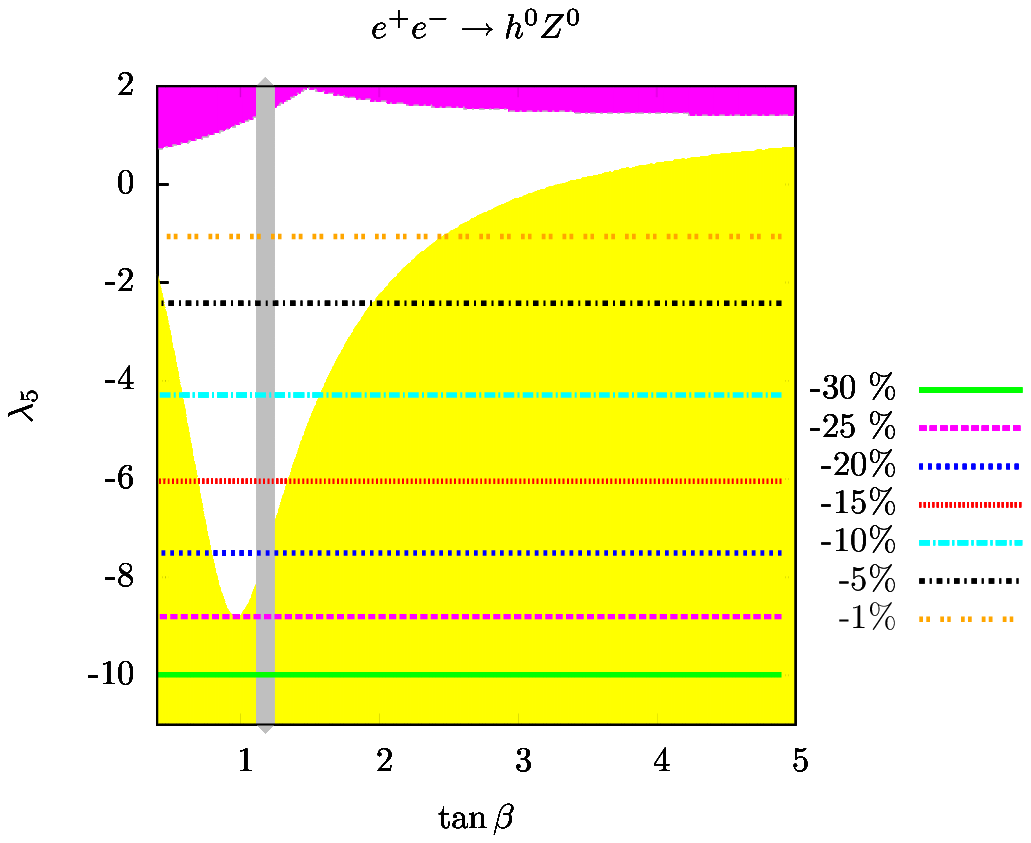}\end{tabular}
\caption{Contour plots for the relative size of the one-loop quantum corrections 
$\delta_r \equiv [\sigma^{(0+1)}-\sigma^{(0)}]/\sigma^{0}$ (in \%) 
to the $\eeAl$ (left panel) and $\eeZl$ (right panel) total rates, as
a function of $\tan\beta$ and $\lambda_5$. We fix
$\alpha = \beta$ (for  $\eeAl$) and $\alpha = \beta-\pi/2$ (for $\eeZl$), 
in which cases their respective born-level couplings maximize. For the Higgs boson masses we use 
Set A from Tab.~\ref{tab:masses}. The linac
center-of-mass energy is taken to be $\sqrt{s} = 500$ GeV. The shaded areas in the 
upper (resp. lower) patches of the $\tan\beta-\lambda_5$ plane are excluded
by unitarity\cite{unitarity} (resp. vacuum stability\cite{vacuum}) bounds. The vertical grey strip
accounts for the lower limit $\tan\beta \simeq 1.18$ stemming from $B_d^0-\bar{B}_d^0$ data \cite{superiso}.} \label{fig:scans}
\end{center}
\end{figure}

Throughout our study we make use of the standard algebraic and numerical
packages {\sc FeynArts, FormCalc} and {\sc LoopTools} \cite{hahn}. 
Updated
experimental constraints (from EW precision data, low-energy flavor-physics 
and the Higgs mass regions ruled out by direct collider
searches), 
as well as theoretical consistency
conditions (perturbativity, unitarity and vacuum stability) are duly taken into account
~\cite{constraints_general,superiso,unitarity,vacuum,2hdmcalc,higgsbounds}. 
For definiteness, we set along two Higgs boson mass benchmark choices A and B,
as quoted in Tab.~\ref{tab:masses}:
\begin{table}[htb]
\begin{center} 
\begin{tabular}{ccccc} 
 & $M_{\hzero}$ [GeV] & $M_{\Hzero}$ [GeV] & $M_{\Azero}$ [GeV] & $M_{\PHiggs^\pm}$ [GeV] \\ \hline
Set A & 130 & 200 & 260 & 300 \\
Set B & 130 & 150 & 200 & 160 
\end{tabular} 
\caption{Choices of Higgs boson masses employed throughout our calculation.}
\label{tab:masses}
\end{center}
\end{table}

\bigskip{} In order to 
carry out the complete one-loop computation
we are entitled to define suitable UV counterterms, in particular for
the renormalization of the Higgs boson masses and wave functions.
These we can express in terms of the Higgs 2-point
functions at order $\mathcal{O}{(\alpha_{ew})}$. 
Conventional on-shell renormalization conditions
-- see e.g. Ref.~\cite{renorm_sm} -- are extended to the 2HDM Higgs sector.
In particular, the relations 

\begin{eqnarray*}
 && \retildehat '_{\PHiggspszero\PHiggspszero}(q^2)\Big]_{q^2=M^2_{\PHiggspszero}} = 0; \quad
 \retildehat_{\PHiggspszero\PZ^0}(q^2)\Big]_{q^2=M^2_{\PHiggspszero}} = 0,
\label{eq:rencon} \end{eqnarray*}

\noindent anchor the wave function
renormalization of the Higgs doublets, and thereby of all 
the physical Higgs fields. The remaining Higgs boson masses, as well as the mixing
angle $\alpha$, are determined via on-shell conditions imposed on their respective
self-energies (including the $\hzero-\Hzero$ kinetic mixing). The parameter
$\tan\beta$ is fixed via Eq.~\eqref{eq:rencon} alonside with one additional condition on the 
Higgs boson tadpoles. An exhaustive account of the renormalization procedure
is available in Ref.~\cite{loop1}.

\subsection{Higgs boson pair production at $\mathcal{O}(\alpha^3_{ew})$: $\eeAlH$}

\begin{figure}[t!]
\begin{center}
 \includegraphics[scale=0.4]{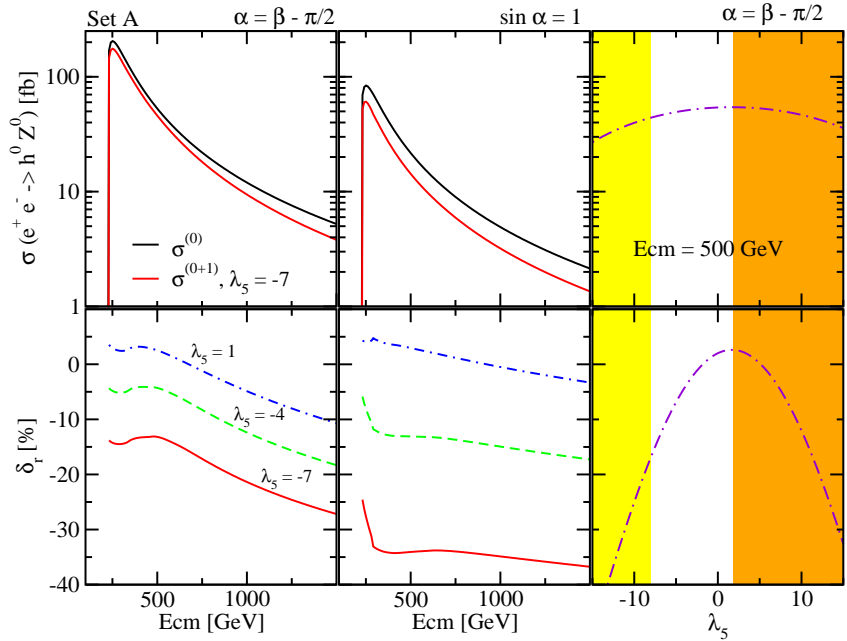}
\caption{Tree-level and loop-corrected cross section [$\sigma(\eeAl)$] (in fb) 
and relative size of the radiative corrections [$\delta_r$] (in \%),
as a function of $\sqrt{s}$ (left, center) and $\lambda_5$ (right), for 
an equivalent setup to that of Fig.~\ref{fig:one}.} \label{fig:two}
\end{center}
\end{figure}

For definiteness, we focus on the light
Higgs mode $[\hzero\Azero]$ and specialize our results for the 
Higgs mass spectrum defined by Set A of Tab.~\ref{tab:masses}. We quantify
our analysis by means of 
i) the Born-level,
[$\sigma^{(0)}$], and 1-loop cross sections, [$\sigma^{(0+1)}$] -- in which we include
the full set of $\mathcal{O}(\alpha^3_{ew})$ corrections, and also the leading
$\mathcal{O}(\alpha^4_{ew})$ ones which arise from the squared of the scattering amplitude \cite{loop1}; 
ii) the relative size of the 1-loop corrections, via
the parameter $\delta_r \equiv [\sigma^{(0+1)}-\sigma^{(0)}]/\sigma^{(0)} $.
The upshot of our findings, as summarized in Fig.~\ref{fig:one}, 
highlights
substantial production rates, which fall roughly in the range of $2-15$ fb
for $\sqrt{s} = 0.5 $ TeV -- this is, up to barely $10^3-10^4$ events per
$500$ fb$^{-1}$; and eventually very large quantum corrections, 
of the order $|\delta\sigma|/\sigma \sim 20-60\,\%$, which can be either positive (for
$\sqrt{s} \simeq 0.5 $ TeV) or negative (
$\sqrt{s} \gtrsim 1 $ TeV) and fairly independent on the
details of the Higgs mass spectrum, the particular value of the
tree-level coupling $[g_{h\Azero\PZ}]$ and the actual type of 2HDM 
under
consideration -- namely, whether we specifically target at the type-I or
II realizations of the 2HDM. The evolution
of $\sigma^{(0+1)}$
and $\delta_r$ as a function of $\sqrt{s}$ for different $\lambda_5$ values
evinces 
how critically the quantum effects depend on the Higgs self-interaction enhancements.
The quadratic dependence on the parameter $\lambda_5$, $\sigma \sim (a - b\lambda_5)^2$, as shown
in the rightmost panel of Fig.~\ref{fig:one}, nicely illustrates the dominance
of the Higgs mediated one-loop diagrams -- these are indeed sensitive to the product
of two triple Higgs self-interactions. As a complementary viewpoint, 
in the left panel of Fig.~\ref{fig:scans} we display the profile
of the radiative corrections $\delta_r$ to the total
cross-section $[\sigma(\hzero\Azero)]$ along the $\tan\beta-\lambda_5$
plane, again for Set A of Higgs boson masses, $\alpha=\beta$ and
a linac center-of-mass energy of $\sqrt{s} = 0.5\,\TeV$. The choice
$\alpha=\beta$ maximizes the tree-level cross section.
Unitarity \cite{unitarity} and vacuum stability limits \cite{vacuum}
(lower and upper shaded areas, respectively)
restrict the largest attainable quantum effects to 
regions with $\tan\beta \simeq 1$ and $|\lambda_5| \sim 5-10 $ ($\lambda_5 < 0$).
The central grey band
depicts the lower limit $\tan\beta \simeq 1.18$ ensuing from $B_d^0
- \bar{B}_d^0$ \cite{superiso}.

\subsection{Associated Higgs/$\PZ^0$-boson production at $\mathcal{O}(\alpha^3_{ew})$: $\eeZlH$}

Again, without loss of generality, we concentrate on the light
Higgs mode $[\hzero\PZ^0]$ and arrange the mass spectrum as in Set A
of Tab.~\ref{tab:masses}. Our results are shown 
in Fig.\ref{fig:two}. In this case we obtain typical
cross sections in the range of $\sigma(\hzero Z^0) \sim
\mathcal{O}(10-100) \,$fb, with very significant (and systematically negative) radiative
corrections (up to order $\delta_r \sim -50\%$),
reaching their maximum again in the 
parameter space regions with $\tan\beta \sim \mathcal{O}(1)$ and $|\lambda_5| \sim
\mathcal{O}(10)$. Such a characteristic pattern of negative quantum
effects we can relate to the dominance of the finite
wave function corrections to the external Higgs boson fields -- this being
the only contribution which retains a quadratic
dependence on $\lambda_{HHH}$ at one loop, as we can also read off the rightmost
panel of Fig.~\ref{fig:two}. The relative size of the quantum effects [$\delta_r$]
and its interplay with the relevant constraints is examined in the right panel 
of Fig.~\ref{fig:scans} as we move across the $\tan\beta-\lambda_5$ plane.
Set A of Higgs boson masses, a fixed value of $\alpha=\beta-\pi/2$ and
a linac center-of-mass energy to $\sqrt{s} = 0.5 \,\TeV$ are employed throughout.
Worth noticing is that the $\delta_r$ isocurves are not responsive to a change of $\tan\beta$.
This follows directly from the analytical structure of all the relevant couplings
for the particular setup $\alpha=\beta-\pi/2$ \cite{loop2} -- which corresponds to the 
decoupling (or SM-like) limit of the 2HDM.

\section{Discussion and closing remarks}
\label{sec:closing}

\begin{table}[tb!]
\begin{center}
\begin{tabular}{|c||c|c|c|}  \hline
Process & $\sigma(\sqrt{s} = 0.5\,\TeV) [\femtobarn]$ & 
$\sigma(\sqrt{s} = 1.0\,\TeV) [\femtobarn]$ & 
$\sigma(\sqrt{s} = 1.5\,\TeV) [\femtobarn]$ \\ \hline \hline
$\hzero\Azero$ &  26.71 ($ \delta_r = 31.32\%$) & 4.07 & 1.27 \\
$\Hzero\PZ^0$ &  19.09 ($\delta_r = -61.56\%$) & 3.73 & 1.47  \\
$\hzero\Hzero\Azero$ & 0.02  & 5.03 & 3.55 \\
$\Hzero\PHiggs^+\PHiggs^-$ & 0.17  & 11.93 & 8.39 \\
$\hzero\hzero + X$ & 1.47  & 17.36 & 38.01 \\ \hline
\end{tabular}
\caption{Compared cross section (in fb) for different associated, pairwise and triple 
Higgs boson production mechanisms at $\mathcal{O}(\alpha^3_{ew},\alpha^4_{ew})$,
for $\tan\beta = 1$, $\alpha = \beta$ and $\lambda_5 = -10$.
The Higgs boson mass spectrum we fix as in Set B of Tab.~\ref{tab:masses}. The relative
size of the one-loop corrections [$\delta_r$] for the Higgs pair and Higgs strahlung
mechanisms is quoted in brackets.} \label{tab:combined}
\end{center}
\end{table}

Higgs boson self-interactions constitute a paradigmatic structure
of extended Higgs sectors of non-supersymmetric nature. The general
(unconstrained) Two-Higgs-Doublet Model is a canonical example of the
latter. Here, the triple and quartic Higgs boson self-interactions
are not subdued by the gauge symmetry. This entails two major consequences,
which are in stark contrast to analogue models, such as e.g. the MSSM: 
i)  the Higgs boson spectrum is fully unconstrained; this is to say, no limitations
on the mass splittings between the physical Higgs boson fields must
be assumed \emph{a priori}; ii) by the very same token, the Higgs
boson self interactions are also fundamentally unrestricted, and hence
may accomodate sizable enhancements. Both features are
tamed in part by stringent theoretical and phenomenological constraints
(unitarity, vacuum stability, electroweak precision and flavor physics) but
nevertheless open up a plethora of rich, and highly distinctive, phenomenological
possibilities. So much so, the analysis of collider 
observables which are sensitive to the Higgs self-interactions, 
either directly or through
quantum corrections, may bring forward instrumental handles to disclose 
non-SUSY vs SUSY multi-doublet Higgs structures.

\smallskip{}
In this contribution we have concisely revisited two particular examples
of Higgs boson production from $\APelectron\Pelectron$ colliders within the 
2HDM context, these are the pairwise production of neutral Higgs
bosons ($\hzero\Azero$/$\Hzero\Azero$) and the Higgs-strahlung channels
($ \hzero\PZ^0/\Hzero\PZ^0 $). We have portrayed their phenomenology
at a future linac and have spelled out the 
features that singularize the 2HDM scenarios with large Higgs boson self-couplings.
Our findings can be outlined as follows:
 
\begin{itemize}
\item{\textbf{Large Higgs boson production rates}, 
in the ballpark of $\sigma_{2h,hZ} \sim \mathcal{O}(10-100)$ fb for 
$\sqrt{s} = 500$ GeV, and yet of few dozens of fb in the TeV-range
center of mass energies -- this would correspond to rates 
of $\mathcal{O}(10^2 - 10^5)$ events for an integrated luminosity of $500 \,\invfb$.}
\item{\textbf{Large quantum effects}, which may reach up to
$\delta_r \equiv [\sigma^{(0+1)}-\sigma^{(0)}]/\sigma^{(0)} \sim \pm 50\%$, 
preferably 
realized within the $\tan\beta \sim \mathcal{O}(1)$ and $|\lambda_5| \sim
\mathcal{O}(10)$ domains of the 2HDM parameter space. These may alternatively lead
to characteristic enhancements (e.g. for $\sigma(2h)$ at $\sqrt{s} \simeq 500 \,\GeV$), 
or suppressions (for $\sigma(hZ)$ and also for $\sigma(2h)$ at larger $\sqrt{s}$)}.

\item{\textbf{A generic phenomenological pattern}, in the sense that the above observations
barely depend on the very choice of Higgs masses nor the type of 
Yukawa couplings to fermions, and they hold for broad regions across the $\tan\beta - \sin\alpha$ plane. }
\end{itemize}

\noindent Interestingly enough, enhancements of the Higgs boson production rates have
 also been put forward in the literature for alternative multi-Higgs production processes, 
such as the 
$\APelectron\Pelectron \to hhh$ \cite{giancarlo} and
$\APelectron\Pelectron \to VV^* \to hh + X$ \cite{neil} channels. 
In this vein, let us consider the
following choice of parameters:
$\tan\beta =1$, $\alpha = \beta$ and $\lambda_5 = -10$, along with Set B
of Higgs boson masses from Tab.~\ref{tab:masses}. This particular configuration
saturates
the unitarity bounds, and thus maximizes the impact of the Higgs boson self-interactions. 
If we now
combine the evaluation of the total production rates for all these different production channels,
we come up with
the cross-correlated set of predictions displayed in Tab.~\ref{tab:combined}.
These results reflect the great complementary of the different multi-Higgs
states at different center-of-mass energies. 
Likewise, 
the correlation of large negative
quantum effects on the Higgs-strahlung channels with the presence of
significant positive (for $\sqrt{s}\lesssim 500$ GeV) and 
negative (for $\sqrt{s}>600$ GeV) quantum effects on the double
Higgs production may eventually constitute an additional hint
at a generic
(unconstrained) 2HDM dynamics.
Notice once more that, in all these cases, the reported pattern of 
signatures crucially relies on the strenght of
the 3H self-couplings. No analogue picture could then be attributed e.g. to the MSSM. 

\smallskip{}
On balance, there is little doubt that a linear collider qualifies as a most cherised tool
to carry to completion the Higgs boson research program.
Owing to its superbly clean environment, a linac facility should enable accurate
measurements of the Higgs boson masses, gauge and Yukawa couplings, as well as of the 
Higgs boson self-interactions themselves. This means, it could provide 
us with the firmest possible grip on the fundamental structure of the EWSB sector. 
In this context, our results 
underline the 
possibilities of
the Higgs boson self-interactions as 
a trademark dynamical feature of the generic 2HDM.
We prove their capabilities to   
rubber-stamp significant -- and highly distinctive -- fingerprints 
on multi-Higgs production processes, 
either at the leading order or through quantum corrections, and conclude that
these might well constitute a pristine window towards non-standard, non-supersymmetric Higgs sectors.

%
%
%
%
%

\section{Acknowledgments}

DLV wishes to thank the LC2012 
local organizing committee and, in particular, Gudrid Moorgat-Pick and Georg
Weiglein, for the hospitality extended to him at DESY and also for
travel support. JS has been supported in part by MEC and
FEDER under project FPA2010-20807, by the Spanish Consolider-Ingenio 2010 program CPAN
CSD2007-00042 and by DIUE/CUR Generalitat de Catalunya under project 2009SGR502.


%
%
%

\newcommand{\JHEP}[3]{ {JHEP} {#1} (#2)  {#3}}
\newcommand{\NPB}[3]{{\sl Nucl. Phys. } {\bf B#1} (#2)  {#3}}
\newcommand{\NPPS}[3]{{\sl Nucl. Phys. Proc. Supp. } {\bf #1} (#2)  {#3}}
\newcommand{\PRD}[3]{{\sl Phys. Rev. } {\bf D#1} (#2)   {#3}}
\newcommand{\PLB}[3]{{\sl Phys. Lett. } {\bf B#1} (#2)  {#3}}
\newcommand{\EPJ}[3]{{\sl Eur. Phys. J } {\bf C#1} (#2)  {#3}}
\newcommand{\PR}[3]{{\sl Phys. Rept. } {\bf #1} (#2)  {#3}}
\newcommand{\RMP}[3]{{\sl Rev. Mod. Phys. } {\bf #1} (#2)  {#3}}
\newcommand{\IJMP}[3]{{\sl Int. J. of Mod. Phys. } {\bf #1} (#2)  {#3}}
\newcommand{\PRL}[3]{{\sl Phys. Rev. Lett. } {\bf #1} (#2) {#3}}
\newcommand{\ZFP}[3]{{\sl Zeitsch. f. Physik } {\bf C#1} (#2)  {#3}}
\newcommand{\MPLA}[3]{{\sl Mod. Phys. Lett. } {\bf A#1} (#2) {#3}}
\newcommand{\JPG}[3]{{\sl J. Phys.} {\bf G#1} (#2)  {#3}}
\newcommand{\JPCF}[3]{{\sl J. Phys. Conf. Ser.} {\bf G#1} (#2)  {#3}}
\newcommand{\FDP}[3]{{\sl Fortsch. Phys.} {\bf G#1} (#2)  {#3}}
\newcommand{\CPC}[3]{{\sl Com. Phys. Comm.} {\bf G#1} (#2)  {#3}}

\begin{footnotesize}



%

\end{footnotesize}


\end{document}